\documentclass[preprint, amsmath, amssymb, aps, prl, superscriptaddress, showpacs, floatfix]{revtex4-1}

\usepackage{graphicx}
\usepackage{dcolumn}
\usepackage{bm}
\usepackage{xcolor}
\usepackage{float}
\usepackage{placeins} 
\usepackage{natbib} 

\begin{document}

\title{Observation of light propagation through a three-dimensional cavity superlattice in a 3D photonic band gap}

\author{Manashee Adhikary}
\affiliation{Complex Photonic Systems (COPS), MESA+ Institute for Nanotechnology, University of Twente, P.O. Box 217, 7500 AE Enschede, The Netherlands} 
\affiliation{present address: Advanced Research Center for Nanolithography (ARCNL), Science Park 106, 1098 XG Amsterdam, The
Netherlands} 

\author{Marek Kozo\v{n}} 
\affiliation{Complex Photonic Systems (COPS), MESA+ Institute for Nanotechnology, University of Twente, P.O. Box 217, 7500 AE Enschede, The Netherlands}
\affiliation{Mathematics of Computational Science (MACS),  MESA+ Institute for Nanotechnology, University of Twente, P.O. Box 217, 7500 AE Enschede, The Netherlands}

\author{Ravitej Uppu} 
\affiliation{Complex Photonic Systems (COPS), MESA+ Institute for Nanotechnology, University of Twente, P.O. Box 217, 7500 AE Enschede, The Netherlands} 
\affiliation{present address: Department of Physics \& Astronomy, University of Iowa, Iowa City, IA 52242, United States.} 

\author{Willem L. Vos}
\affiliation{Complex Photonic Systems (COPS), MESA+ Institute for Nanotechnology, University of Twente, P.O. Box 217, 7500 AE Enschede, The Netherlands} 

\begin{abstract}
We experimentally investigate unusual light propagation inside a three-dimensional (3D) superlattice of resonant cavities that are confined within a 3D photonic band gap. 
Therefore, we fabricated 3D diamond-like photonic crystals from silicon with a broad 3D band gap in the near-infrared and doped them with a periodic array of point defects. 
In position-resolved reflectivity and scattering microscopy, we observe narrow spectral features that match well with superlattice bands in band structures computed with the plane wave expansion. 
The cavities are coupled in all three dimensions when they are closely spaced ($a_{SL} < 4a$), and uncoupled when they are further apart ($a_{\mathrm{SL}} > 4a$). 
The superlattice bands correspond to light that hops in high-symmetry 
directions in 3D - so-called ``Cartesian light'' - that opens applications in 3D photonic networks, 3D Anderson localization of light, and future 3D quantum photonic networks. 
\end{abstract}

\maketitle

\begin{figure}[htbp]
\centering
\includegraphics[width = 0.4\columnwidth]{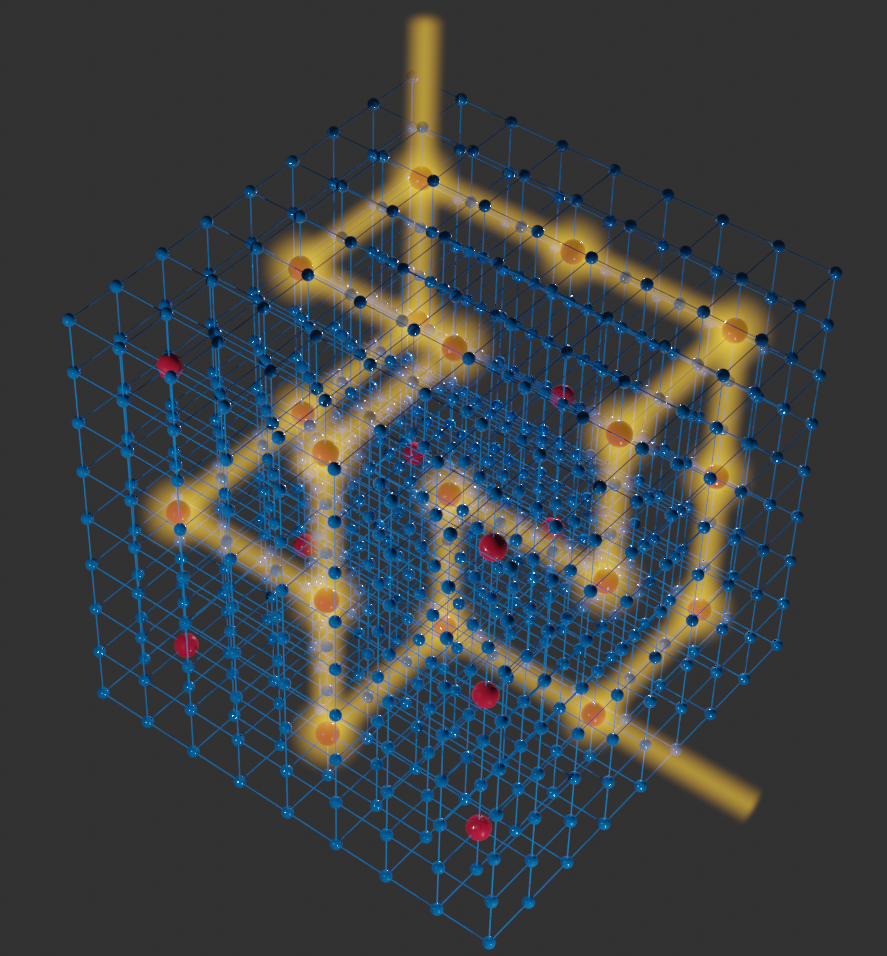}
    \caption{Schematic of a few possible paths for waves (yellow) hopping in Cartesian 
    directions in a 3D cavity superlattice. 
    At the end of their Cartesian journey, waves may exit from a different crystal surface (top) than where they entered (right). 
    Each cavity is shown as a red sphere and the surrounding (photonic band gap) crystal as the 3D mesh with blue spheres as unit cells. 
    }
    \label{fig:cartoon}
\end{figure}

Many fruitful analogies exist between waves such as light or sound that propagate through mesoscopic photonic or phononic metamaterials and the well-known propagation of elementary excitations in atomic crystals such as phonons, electron and spin waves~\cite{vanHaeringen1990Book, Soukoulis2001Book, Sheng2006book, Akkermans2007book, Kruglyak2010JPD, Ghulinyan2015Book}. 
The physics of both classes of waves is governed by multiple scattering from mesoscopic potentials and concomitant interferences. 
A peculiar class of transport is discretized wave transport with hopping in all three dimensions on superlattices, which is known for phonons, electron, and spin waves, but not for electromagnetic waves and light. 
A superlattice is a periodic arrangement of a supercell that consists itself of multiple unit cells of an underlying crystal structure~\cite{Bragg1934RSL, Bethe1935RSL}; its applications include photovoltaics since the absorption is enhanced by intermediate electron bands~\cite{Li1992APL,Luque1997PRL,TableroSEMSC2005,SullivanJAP2013,Liu2018SR}. 

In one dimension (1D) discretized transport of light is well known to occur in coupled resonator optical waveguides (CROW), where an optical resonator is so close to its neighbors that light hops from one resonator to a neighbor due to evanescent coupling~\cite{Yariv1999OL, Bayindir2000PRB, Altug2005OpEx}. 
A well-known feature is that the bandwidth of the resulting defect band is \textit{proportional} to the hopping rate, as is known from atomic solid state physics~\cite{Ashcroft1976book}. 
In a recent theoretical study of our team, we identified a major distinction with 3D wave transport in 3D superlattices~\cite{Hack2019PRB}: in 3D the bandwidth of the defect band is \textit{not proportional} to the hopping rates, but determined by intricate interferences between hopping in different high-symmetry directions. 

Therefore, we perform in this paper an original experimental study of a 3D photonic band gap cavity superlattice, with a band of photonic states. 
Metaphorically, we cross the bridge from 
metamaterial physics to atomic condensed matter physics, using an innovative 3D nanophotonic system: 
we fabricate 3D photonic crystals with a complete 3D photonic band gap~\cite{Yablonovitch1987PRL, John1987PRL, Joannopoulos2008Book} that are doped by judiciously placing resonating cavities in a periodic fashion~\cite{Hack2019PRB, Kozon2022PRL}, see Fig.~\ref{fig:cartoon}. 
Since each cavity confines light with a photon energy within the band gap, photons can only propagate by effectively hopping from one cavity to a neighboring cavity on discrete lattice positions in 3D. 
We observe resonant scattering peaks that match well with theoretical defect bands. 
When we excite a defect band on resonance, we observe states that spatially extend over multiple superlattice positions $\mathbf{R}$. 
Conversely, if we excite off-resonance, the observed states are localized on single cavity positions. 
Therefore, light hops in 3D in high-symmetry 
directions in a crystal, known as Cartesian light, see Fig.~\ref{fig:cartoon}, which opens avenues to applications in 3D classical and quantum photonic networks~\cite{Crespi2016Natcomm,Hoch2022npjQI,Moughames2020Optica}, and 3D Anderson localization of light~\cite{Skipetrov2014prl}. 

\begin{figure}[htbp]
    \centering
    \includegraphics[width = 0.6\columnwidth]{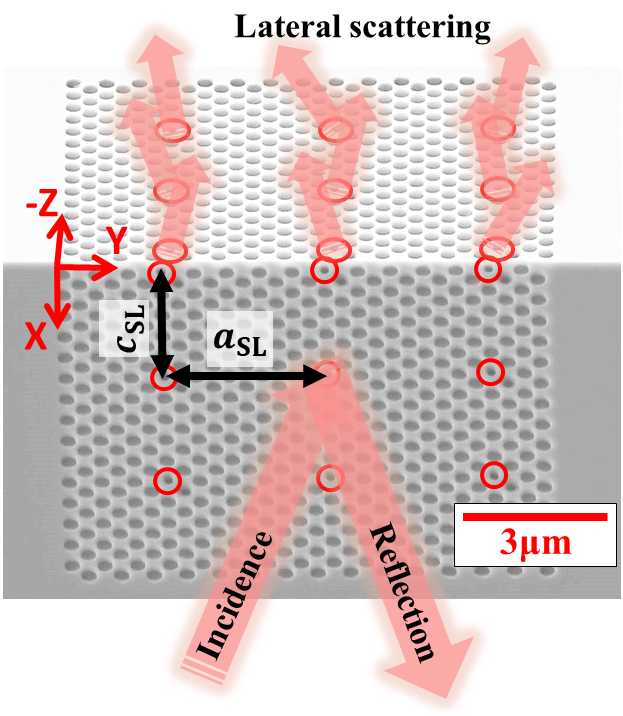}
    \caption{
    Scanning electron micrograph of a 3D cavity superlattice embedded in a 3D photonic band gap crystal made of silicon, viewed in the (X,Z)-direction. 
    The inverse woodpile crystal has lattice constants $a=0.68~\mathrm{\mu m}$ in the Y-direction and $c=0.48~\mathrm{\mu m}$ in the X, Z-directions. 
    The designed pore radius is $r = 0.16~\mathrm{\mu m}$ for the main nanopores and $r'=r/2=0.08~\mathrm{\mu m}$ for the defect pores.
    The superlattice of the point defects has the unit cell with dimensions $5c, 5a, 5c$.
    The excess silicon at the intersections of these defect pores inside the material form point defects, constituting the superlattice of 3D cavities. 
}
    \label{fig:SEM}
\end{figure}

The 3D photonic band gap crystals that we study have a diamond-like inverse woodpile structure~\cite{Ho1994SSC} that consist of two arrays of nanopores with radius $r$ running in the perpendicular $\mathrm{X}$ and $\mathrm{Z}$ directions in the high-index silicon backbone \footnote{Each array is rectangular with lattice constants $a = 680$ nm and $c = 480$ nm. By taking the ratio to be $a/c=\sqrt{2}$ the 3D photonic crystal has a cubic crystal structure that is diamond-like~\cite{Ho1994SSC}.}, as illustrated on YouTube~\cite{COPS2012youtube}. 
We employ deep reactive ion etching of silicon through a 3D etch mask on the edge of a long beam~\cite{Tjerkstra2011JVSTB, Grishina2015Nanotech, Grishina20173DNanophotonics}.
Silicon inverse woodpile crystals have a broad and robust 3D photonic band gap~\cite{Hillebrand2003JAP, Maldovan2004NM, Woldering2009JAP, Adhikary2020OpEx} that is well suited to confine embedded cavities and shield them from the surrounding vacuum. 
To define a 3D cavity in an inverse woodpile structure, we design two perpendicular and proximal defect nanopores to have smaller radii $r^{\prime}$ (see Supplementary)~\cite{Woldering2014PRB}. 
Near the intersection of two defect pores, the excess silicon backbone results in a donor-like cavity~\cite{Joannopoulos2008Book}. 
When multiple cavities are spaced periodically in the crystal, coupling between the cavities results in a 3D cavity superlattice that sustains Cartesian light~\cite{Hack2019PRB}. 

We fabricated two types of superlattices with different spacings between the cavities, see Fig.~\ref{fig:SEM} for a scanning electron microscopy (SEM) image of an actual crystal. 
Since the defect pores are separated by five lattice constants (called SL5), the superlattice has lattice constants $a_{SL} = 5a$ in the Y and $c_{SL} = 5c$ in the X- and Z-directions. 
There are $3^2 = 9$ defect pores on each crystal surface, hence up to $3 \times 3^2 = 27$ pore crossings and concomitant cavities in the superlattice. 
We also studied superlattices with lattice spacings $ a_{SL} = 3a $ and $ c_{SL} = 3c $, called SL3, with up to $5^3 = 125$ cavities. 

\begin{figure}
    \centering
    \includegraphics[width = \textwidth]{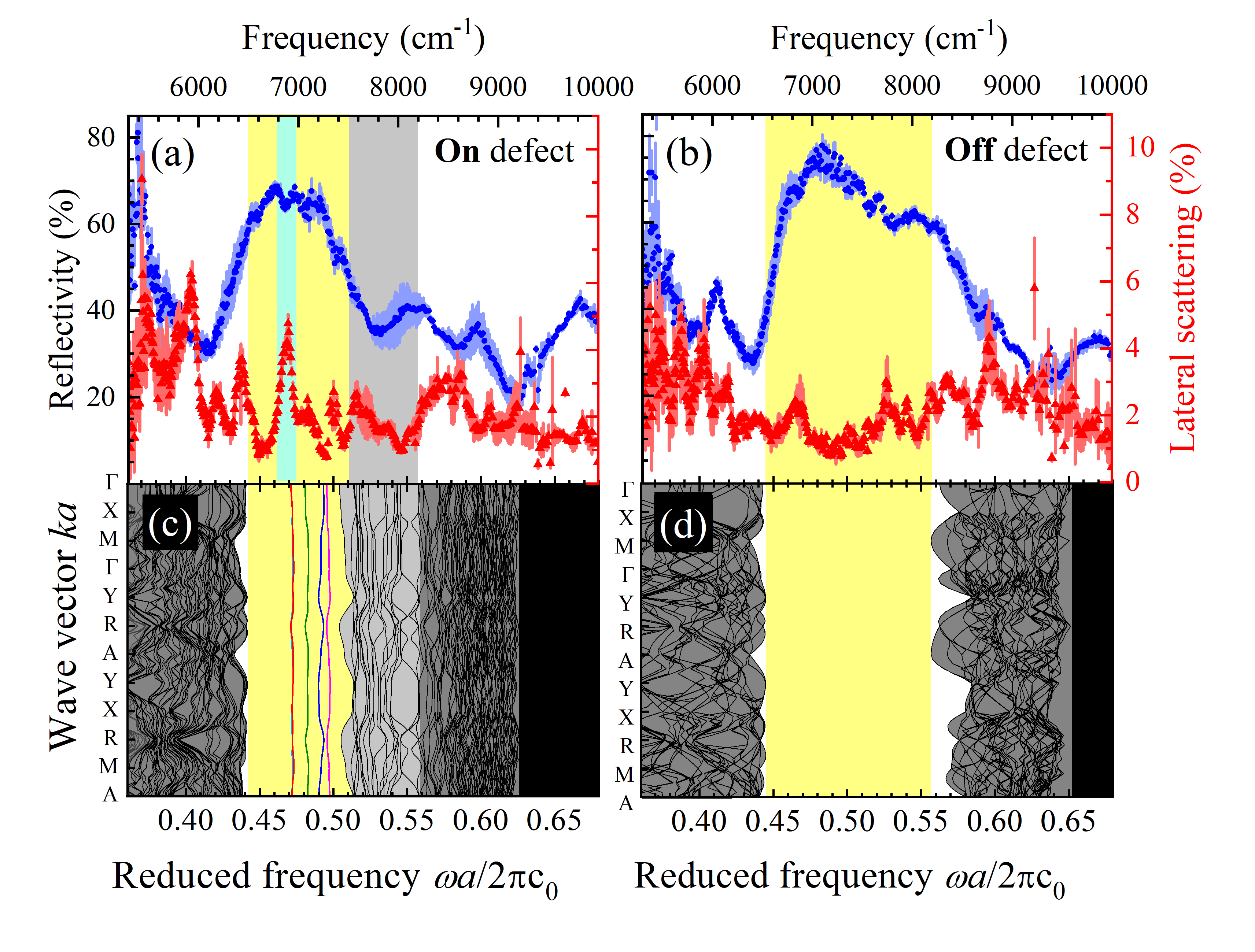}
    \caption{
    (a,b) Measured reflectivity (blue dots) and lateral scattering (red triangles, right ordinate) spectra of the cavity superlattice SL5 on a defect and in between two defects ($X = 2.5~\mathrm{\mu m}$), collected with s-polarization (E-field $\perp$ X-directed pores). 
    The broad reflectivity peaks correspond to the band gaps of the crystal at the particular location, as indicated by the yellow regions that match very well with theory. 
    In (a), the lateral scattering peak at $6900~\mathrm{cm^{-1}}$ within the band gap (cyan highlight) corresponds to a reflectivity trough and a Cartesian band in the band structures. 
    (c) Photonic band structure of a cavity superlattice in an inverse woodpile crystal, rotated for comparison with measured spectra, with frequency and wave vector reduced by the lattice parameter $a$, and with $\epsilon = 12.1$ typical of Si. 
    Crystal pores have a relative radius $r/a = 0.22$ and defect pores $r' = r/2$. 
    The yellow region between $a/\lambda = 0.44$ and $0.502$ is the 3D photonic band gap of the superlattice crystal, and the allowed bands outside the band gap are shown as dark grey areas. 
    The cavity superlattice sustains two types of bands inside the band gap: flat bands typical of Cartesian light in the lower half of the gap
    (colored bands), and other bands in the upper half of the gap (black bands).
    (d) Photonic band structure of an inverse woodpile photonic crystal with $r/a=0.22$ without cavities.
    The yellow region between $a/\lambda = 0.44$ and $0.56$ is the unperturbed 3D band gap. 
    }
    \label{fig:4panel_BSandSpectra}
\end{figure}
To study the cavity superlattices, we built a confocal microscopy setup (see Supplementary Fig.~\ref{fig:fullsetup}) that simultaneously collects both broadband specular reflectivity to probe the band gap and specular cavity scattering, and lateral scattering by the cavities into non-specular directions, see Fig.~\ref{fig:cartoon}. 
Our setup has a frequency range $5300$ to $11000~\rm{cm}^{-1}$ (or $910 < \lambda < 1880$~nm) in the near-infrared that is compatible with silicon nanophotonics. 
The diffraction-limited focus of diameter $1.5~\mathrm{\mu m}$ is smaller than the $3.4~\mu$m distance between neighboring cavities. 
Therefore, in spatial scans across the superlattice surface, we spatially probe both \textbf{on} and \textbf{in between} the cavities. 

Fig.~\ref{fig:4panel_BSandSpectra} shows representative reflectivity and lateral scattering spectra for two main situations, namely \textbf{on} a cavity (a,c), and in between (b,d)
\footnote{In Fig. \ref{fig:4panel_BSandSpectra}(a) the incident focus is close to the middle defect pore on the XY surface, and in Fig. \ref{fig:4panel_BSandSpectra}(b) the incident focus is in between the middle and the rightmost defect pore on the XY surface at $X = 2.5~\mathrm{\mu m}$.}.
Since the $3.4~\mu$m distance between two cavities is much greater than the Bragg length ($L_{B} = 0.2~\mu$m~\cite{Devashish2017PRB}), spectra collected in between cavities effectively correspond to probing the unperturbed crystal.
In both cases we observe an intense and broad reflectivity peak that corresponds to the 3D band gap of the crystal, see also Ref.~\cite{Adhikary2020OpEx}. 
While common lore has it that an increased number of defects in a photonic crystal adversely influences a band gap, ultimately even closing it~\cite{Lidorikis2000PRB, Li2000PRB, Koenderink2005PRB}, we see here that the introduction of a superlattice of defects does \textit{not} close the band gap, hence the superlattice modes identified below are indeed confined in a 3D band gap. 

\begin{figure}
    \centering
    \includegraphics[width = 0.6\columnwidth]{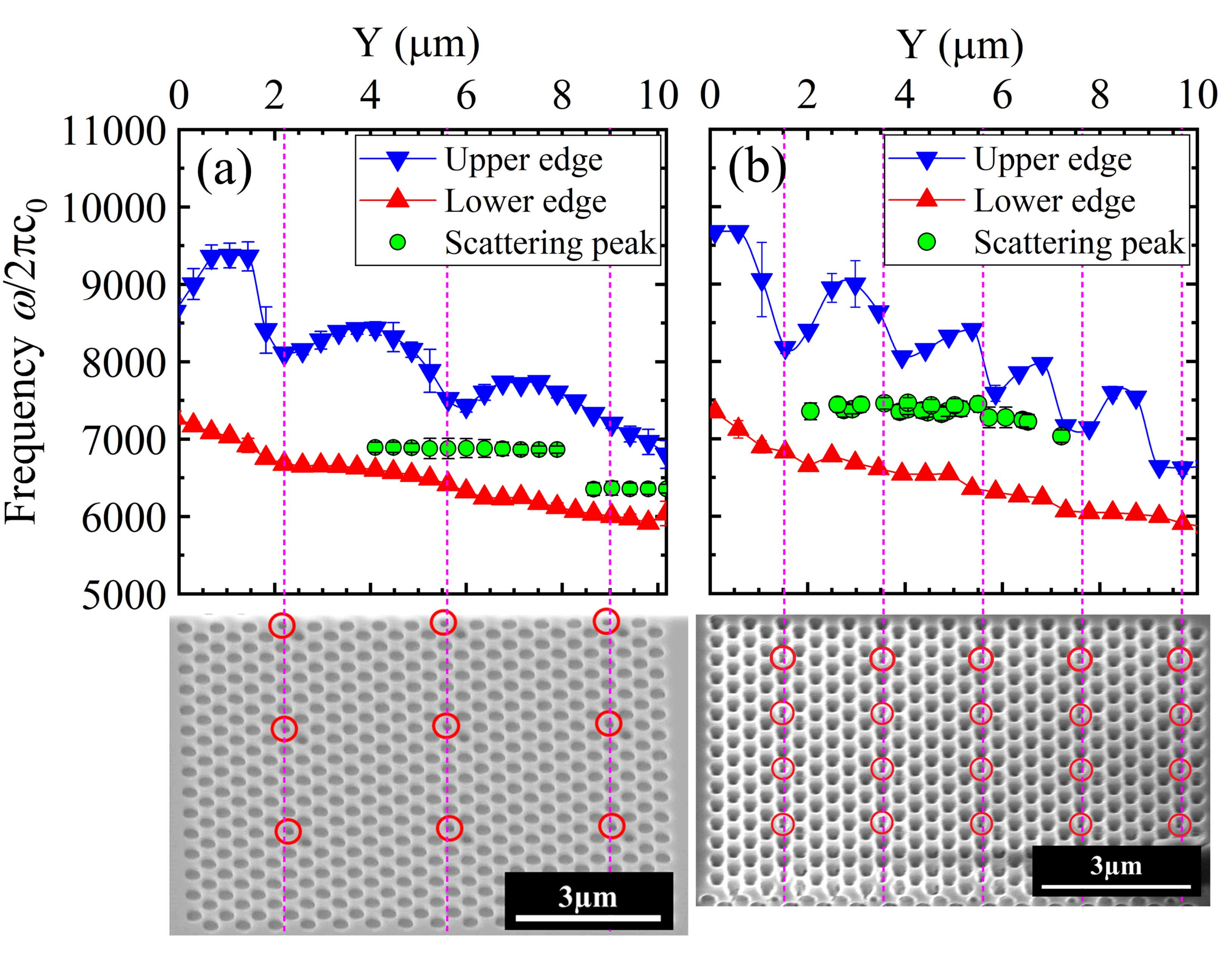}
    \caption{
    Lower edges (red upright triangles) and upper edges (blue downward triangles) of the position-dependent photonic gap versus $ Y $, obtained from reflectivity measurements on two crystals with a cavity superlattice with different lattice spacings. 
    Frequency of the peak inside the stopband obtained from lateral scattering measurements are shown by green circles on each panel.  
    Below each plot the SEM image of the surface of the crystals are stitched with same scale for reference for the reader. 
    The magenta dashed lines trace the positions of the defect pores. 
    }
    \label{fig:bandedges}
\end{figure}

To identify reflectivity features in Fig.~\ref{fig:4panel_BSandSpectra}(a), we compare the measurements with the band structures in Fig.~\ref{fig:4panel_BSandSpectra}(c), computed for a cavity superlattice with pore radii corresponding to those obtained at the probing location using the method 
\footnote{This procedure is necessary since the nanopore radii vary across the sample surface and in depth. 
Therefore, at every probed location there is effectively a `local' pore radius. 
In Ref.~\cite{Adhikary2020OpEx} it is shown that the lower band edge combined with the known gap map~\cite{Woldering2009JAP} gives a reliable estimate of the local pore radius} 
from Ref.~\cite{Adhikary2020OpEx}. 
The band structure shows that the cavity superlattice sustains two types of bands inside the 3D band gap, namely flat defect bands typical of Cartesian light, and other dispersive bands in the upper half of the band gap~\cite{Kozon2022PRL}. 
Due to the appearance of the dispersive bands, the band gap of the cavity superlattice structure shown in Fig.~\ref{fig:4panel_BSandSpectra}(a) is markedly narrower than the unperturbed band gap shown in Fig.~\ref{fig:4panel_BSandSpectra}(b). 
This is clearly observed in the superlattice spectra, where a broad trough near $7800~\mathrm{cm^{-1}}$ coincides with the dispersive bands in the upper half of the original band gap, and thus narrows the superlattice band gap~\footnote{The modulation of the upper edge of the measured gap versus position is clearly seen in Fig.~\ref{fig:bandedges}, where the gap narrows at every defect pore and widens in between.}. 
The center frequency and the width of the measured reflectivity band gap of the superlattice (see Figs.~\ref{fig:4panel_BSandSpectra}(a,b)) agree very well with the theoretical band gap (see Figs.~\ref{fig:4panel_BSandSpectra}(c,d), respectively), which highlights that the fabricated nanostructure and its optical functionality match closely with the designed structure and the intended optical behavior. 
We also observe that upon focusing the incident light on the cavity, the maximum reflectivity ($\mathrm{R_{max}} = 68 \%$, see Fig.~\ref{fig:4panel_BSandSpectra}(a)) is less than when focusing away from the cavities ($\mathrm{R_{max}} = 74 \%$, see Fig.~\ref{fig:4panel_BSandSpectra}(b)), which is reasonable since the light incident on cavities is also scattered non-resonantly, thereby reducing the specular reflectivity. 

When focusing on a cavity, the lateral scattering spectrum shows a distinct peak at $6900~\mathrm{cm^{-1}}$ within the band gap, see Fig.~\ref{fig:4panel_BSandSpectra}(a). 
The peak represents light scattered from incident wave vectors $\mathbf{k_{\rm in}}$ to outgoing wave vectors $\mathbf{k_{\rm out}}$ in completely different directions than either the incident ones $(\mathbf{k_{\rm out}} \neq \mathbf{k_{\rm in}})$ or their Bragg diffracted counterparts $(\mathbf{k_{\rm out}} \neq \mathbf{k_{\rm in}} + \mathbf{G})$. 
Therefore, we expect less light to be reflected and, indeed, the reflectivity spectrum reveals a corresponding trough in the same frequency band. 
Moreover, the scattering peak and the reflectivity trough match well with a Cartesian superlattice band in the band structures, see Fig.~\ref{fig:4panel_BSandSpectra}(b). 
We verified that the scattering peak differs from random speckle that is also observed, see \ref{fig:LSspectra} in supplementary material. 
The observation of resonant features both in specular reflectivity and in scattering confirms that the confinement of light in the superlattice is truly a 3D phenomenon, as opposed to 1D confinement in a Fabry-P\'erot cavity that would be apparent in reflectivity but not in lateral scattering. 
We emphasize that the peak only appears when the incident light is focused onto a defect; when the incident light is focused in between defects the peak is absent, see 
Fig.~\ref{fig:4panel_BSandSpectra}(b)). 
Hence we conclude that the scattering peak corresponds to a Cartesian superlattice band \footnote{
The sum of the reflectivity and the lateral scattering are less than $100 \%$ since our setup can only access part of the light scattered in the $-\mathrm{X}$-direction into air, whereas light scattered in the $(+\mathrm{X}, \pm \mathrm{Y})$-directions escapes undetected. 
Since these scattered contributions propagate in the high-index Si substrate, it is conceivable that their contributions are considerably larger than the detected fraction in air, that may in turn also be attenuated by Fresnel reflectivity at the crystal-air interface.}.

\begin{figure}
    \centering
    \includegraphics[width = 0.6\columnwidth]{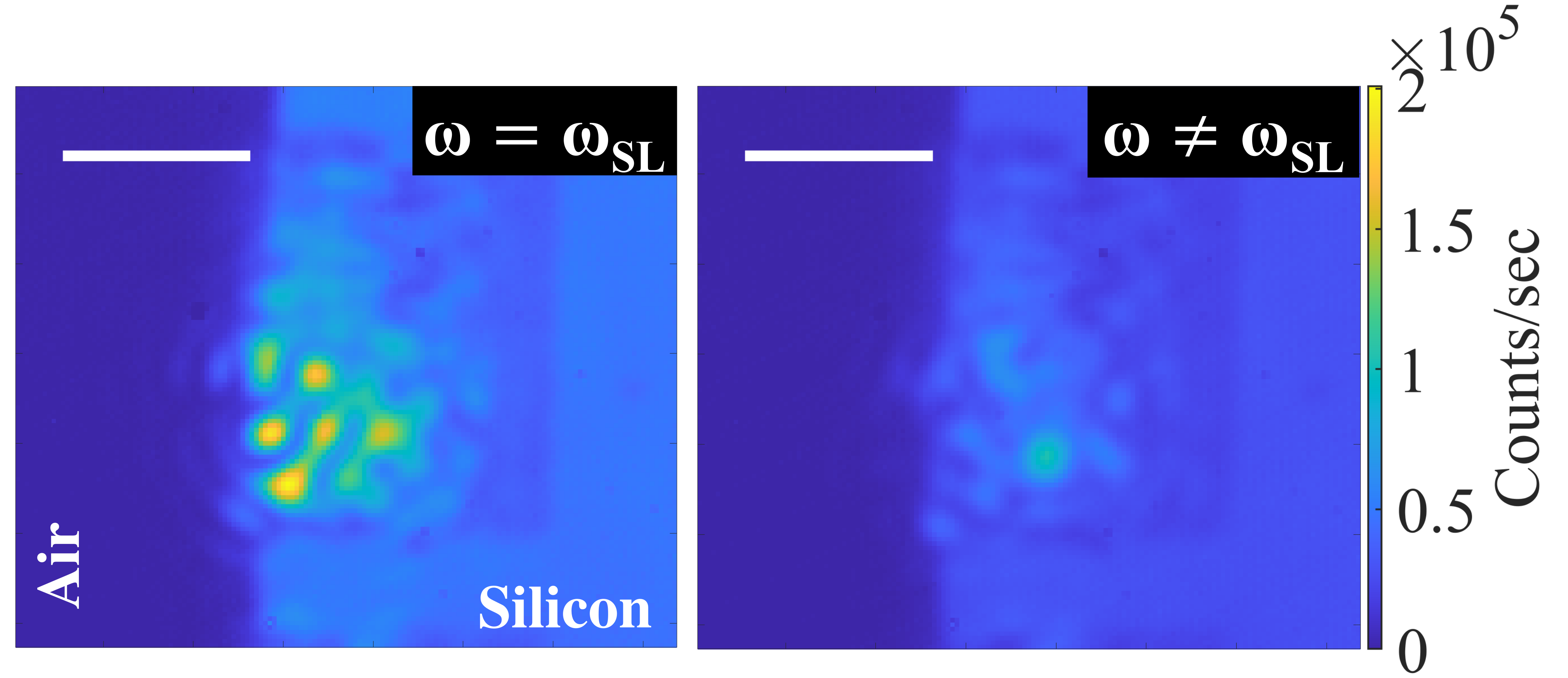}
    \caption{Camera images of the front surface of superlattice SL3 for cross-polarized light taken at two different frequencies. 
    Left: at $\omega =
    7386~\mathrm{cm^{-1}} = \omega_{\rm{SL}}$, \emph{i.e.} at the center of the superlattice peak. 
    Right: at $7174~\mathrm{cm^{-1}}$, outside the superlattice peak.
    The surface of the crystal is illuminated by a separate LED to reveal rectangular XY crystal surface.
    On top left of each figure, the white bar on the top left indicates scale of $5~\mu$m
    }
    \label{fig:speckleimage}
\end{figure}
{\color{black}
To visualize the hopping of light in the superlattice, we capture images of the front surface of SL3 while varying the frequency. 
We detect crossed polarized light to suppress single scattering from the sample surface and select light that has multiply scattered inside the crystal.
Hence, we expect to detect both light scattered from the cavities and from unavoidable disorder. 
Fig.~\ref{fig:speckleimage} presents images for two frequencies $\omega$, at the center of the superlattice peak $\omega = \omega_{\rm{SL}}$ ($=  7386~\rm{cm^{-1}} $), and away $\omega \neq \omega_{\rm{SL}}$ ($= 7174~\rm{cm^{-1}}$), within the unperturbed original band gap. 
In both cases, we see a speckle pattern due to multiple scattering by randomness. 
A remarkable observation is that at $\omega = \omega_{SL}$, the light spreads over a large area, much larger than the incident spot. 
Intensity maxima near the defect pores are observed, as shown in the left image in Fig. \ref{fig:speckleimage}.
In the X-direction, the three intensity maxima are equally spaced, with an average distance of $ 1.50 \pm 0.06~\mu$m. 
This agree very well with the distance between cavities in the X-direction of $ c_{\rm{SL}} = 3c = 1.44~\mu\rm{m}$. 
The good agreement proves that in SL3, the superlattice frequency hops over at least 3 neighboring cavities, which firmly establishes that we observe light hopping in a 3D superlattice. 
}

Due to an inadvertent nanofabrication inhomogeneity of the underlying band gap crystals, we find from the position-dependent measurement that the band gaps shift across the crystals. 
Fig.~\ref{fig:bandedges}(a,b) shows that the band edges shift to lower frequencies while the superlattices are scanned from $Y = 0$ to $10~\mu$m. 
In superlattice SL5 (Fig.~\ref{fig:bandedges}(a)), we observe near the 2nd defect pore a scattering resonance centered at $6896~\mathrm{cm^{-1}}$. 
Scanning further to the 3rd defect pore, the scattering resonance is replaced by another one at a lower frequency $6358~\mathrm{cm^{-1}}$. 
Since the peaks differ for the different defects, the cavities are uncoupled in the Y-direction. 
Strikingly, in superlattice SL3 the scattering peak has the same center frequency ($7386~\mathrm{cm^{-1}}$) when scanned across multiple cavities, see Fig.~\ref{fig:bandedges}(b). 
Therefore, multiple cavities are coupled, hence the peak represents a \textit{bona fide} superlattice band typical of Cartesian light that hops through the superlattice~\cite{Hack2019PRB}. 
To the best of our knowledge, this is the first experimental observation of 3D discretized light transport in a cavity superlattice.

In the pursuit of 3D Anderson localization of light in 3D~\cite{Skipetrov2014prl}, it has been suggested that an interesting approach is the analogy with Anderson's original work~\cite{Anderson1958PR}, namely impurities in crystals. 
If we conceive of cavities as impurities in a photonic band gap, the variation of the individual cavities' resonance frequency and inter-cavity hopping rates correspond to diagonal and off-diagonal disorder in the Anderson model, respectively. 

For future applications as 3D photonic network with potentially more degrees of control than ready-made networks~\cite{Tajiri2019Optica}, we propose to consider adding phase control to the waves hopping in 3D from cavity to cavity, \textit{e.g.}, by wavefront shaping tools~\cite{Rotter2017RMP}. 
Hence, light can take a subset of all possible paths (cf. Fig.~\ref{fig:cartoon}), and thereby, one obtains control on how (which paths) and where (which cavity) the light travels. 
Since photonic control is nowadays feasible down to single photons~\cite{uppu2020SciAdv}, the notion above could find application in 3D quantum networks with significant network size scaling advantages compared to conventional 2D networks~\cite{Moughames2020Optica,Hoch2022npjQI}. 

\section*{Acknowledgments}\label{sec:acknowledgments}
We thank Cock Harteveld for technical help and sample preparations, Chris Toebes for contributions to the initial experiments, Ad Lagendijk for his valuable feedback, and Arie den Boef, Patrick Tinnemans, Vahid Bastani, Wim Coene, and Scott Middlebrooks (ASML) for helpful discussions. 
This research is supported by NWO-TTW Perspectief program P15-36 ``Free-form scattering optics" (FFSO).

\bibliography{refs_3D_cavity_superlattice}
\clearpage
\clearpage

\pagebreak

\onecolumngrid
\begin{center}
  \textbf{\large Observation of light propagation through a three-dimensional cavity superlattice in a 3D photonic band gap\\
  (Supplementary Material)}\\[.2cm]
\author{Manashee Adhikary}
\affiliation{Complex Photonic Systems (COPS), MESA+ Institute for Nanotechnology, University of Twente, P.O. Box 217, 7500 AE Enschede, The Netherlands} 
\affiliation{present address: Advanced Research Center for Nanolithography (ARCNL), Science Park 106, 1098 XG Amsterdam, The
Netherlands} 

\author{Marek Kozo\v{n}} 
\affiliation{Complex Photonic Systems (COPS), MESA+ Institute for Nanotechnology, University of Twente, P.O. Box 217, 7500 AE Enschede, The Netherlands}
\affiliation{Mathematics of Computational Science (MACS),  MESA+ Institute for Nanotechnology, University of Twente, P.O. Box 217, 7500 AE Enschede, The Netherlands}

\author{Ravitej Uppu} 
\affiliation{Complex Photonic Systems (COPS), MESA+ Institute for Nanotechnology, University of Twente, P.O. Box 217, 7500 AE Enschede, The Netherlands} 
\affiliation{present address: Department of Physics \& Astronomy, University of Iowa, Iowa City, IA 52242, United States.} 

\author{Willem L. Vos}
\affiliation{Complex Photonic Systems (COPS), MESA+ Institute for Nanotechnology, University of Twente, P.O. Box 217, 7500 AE Enschede, The Netherlands} 
\end{center}

\setcounter{equation}{0}
\setcounter{figure}{0}
\setcounter{table}{0}
\setcounter{page}{1}
\renewcommand{\theequation}{S\arabic{equation}}
\renewcommand{\thefigure}{S\arabic{figure}}

In this Supplementary, we provide more details on the structure of the inverse woodpile crystals, on the definition of the cavities. 
We describe the optical setup in detail, as well as a number of experimental procedures. 
We show additional experimental results, notably the reproducibility of the lateral scattering to distinguish superlattice peaks from random speckle. 

\subsection{Inverse woodpile photonic crystals and cavity design}

\begin{figure}[htbp]
\includegraphics[width=0.4\columnwidth]{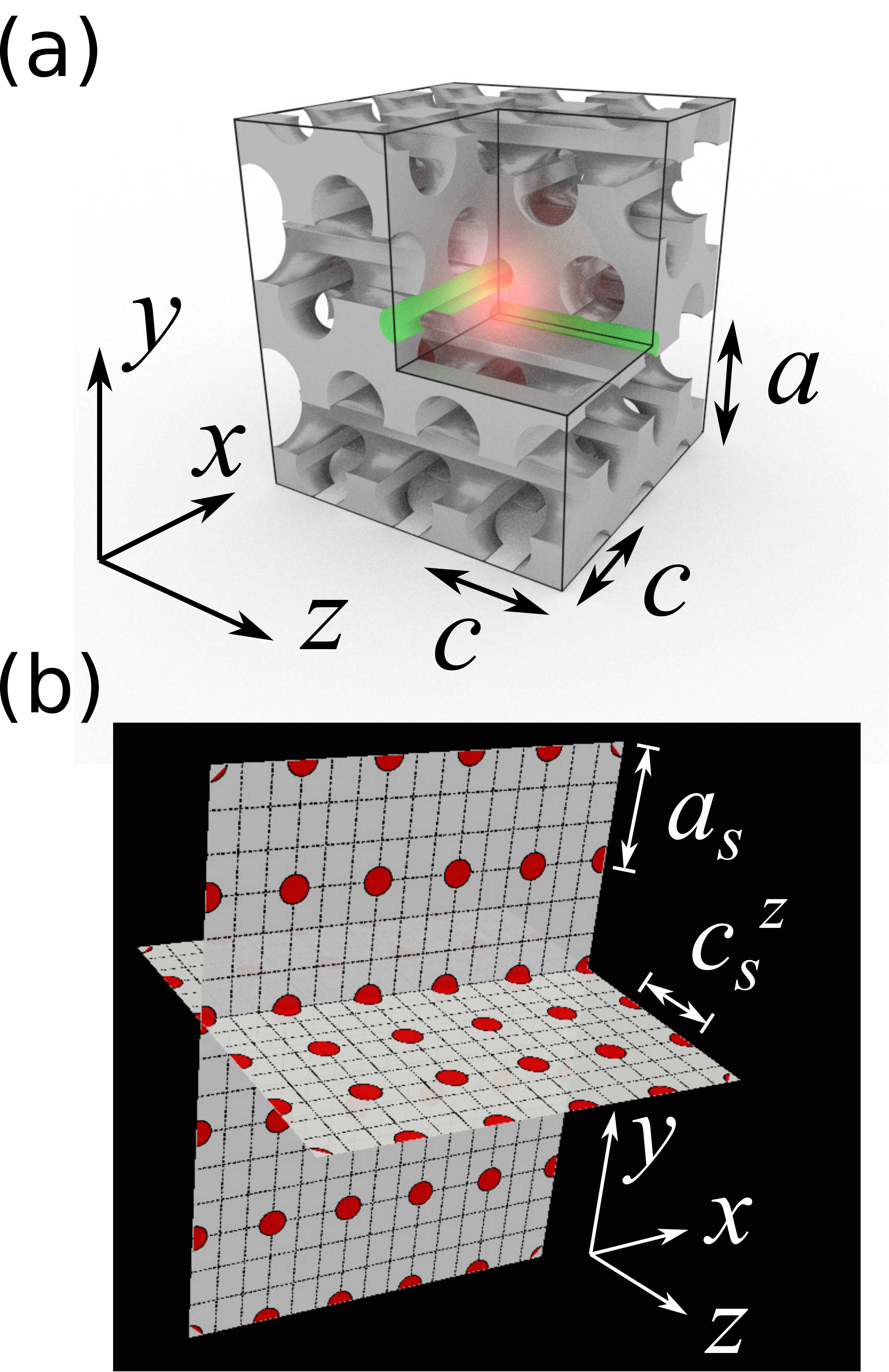}
\caption{
(a) 
Design of a single cavity in an inverse woodpile photonic band gap crystal shown in a cut-out of a $M_x \times M_y \times M_z = 3\times2\times3$ supercell that is surrounded by boxed lines. 
The high-index backbone is shown in gray. 
Two proximal smaller defect pores are indicated in green, and the cavity region is highlighted as the bright region at the center. 
The tetragonal lattice parameters $a$ and $c$ are shown, as well as the $x,y,z$ coordinate system. 
(b) $(x,z)$ and $(x,y)$ cross sections through a 3D superlattice of resonant cavities, with red circles indicating cavities and dashed rectangles representing unit cells of the underlying inverse woodpile crystal structure (see (a)). 
The lattice parameters ($c_s^{x}$, $a_s$, $c_s^{z}$) of the superlattice are shown, as well as the $x,y,z$ coordinate system. 
}
\label{fig:crystalOfCavities}
\end{figure}
\begin{figure}[htbp]
\includegraphics[width=0.5\columnwidth]{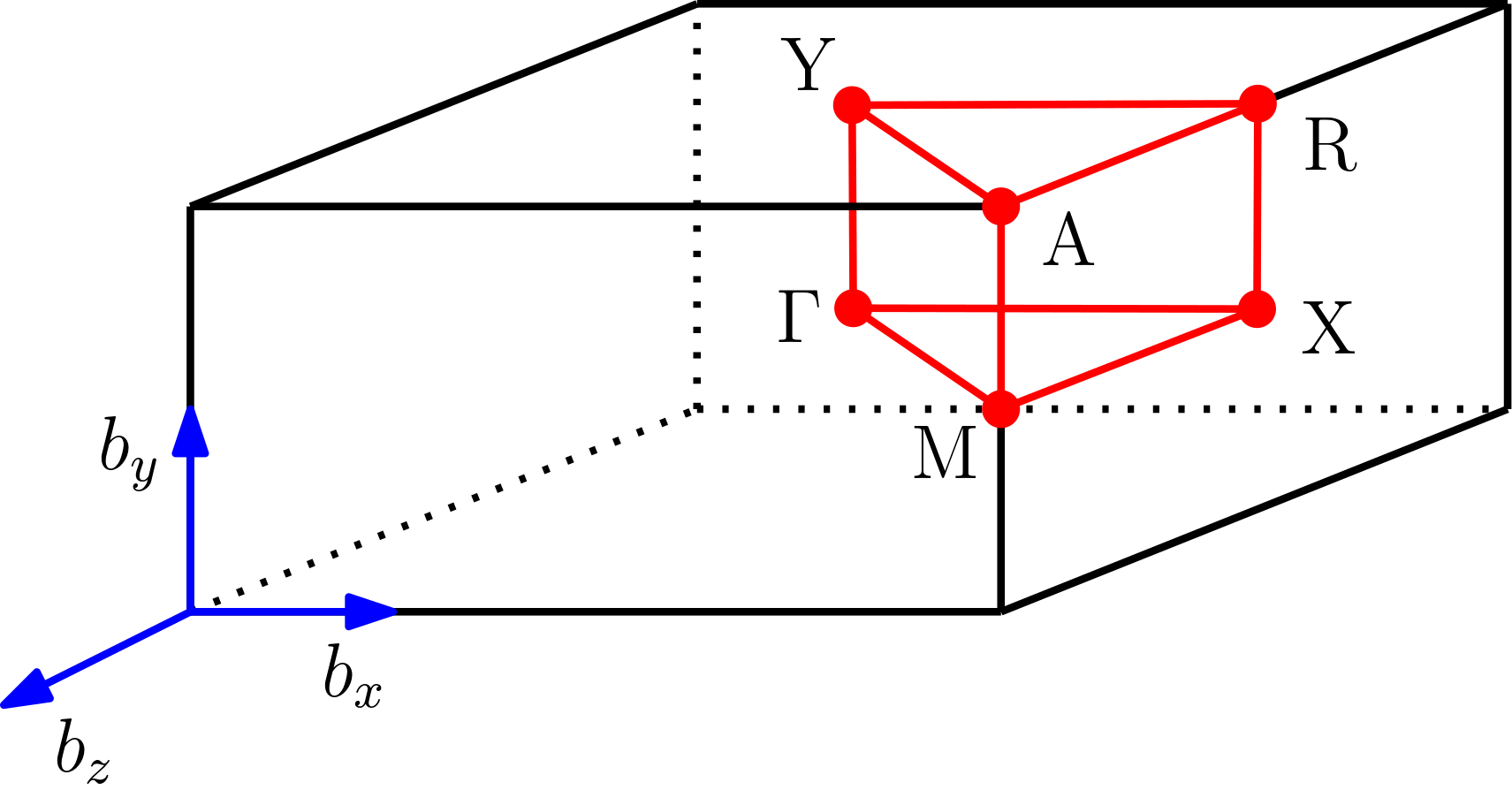}
\caption{ 
Brillouin zone of a tetragonal unit cell (black) with the corresponding high-symmetry path (red). The Cartesian basis vectors (blue) are denoted by $b_x, b_y, b_z$. For more details, see Ref.~\cite{Setyawan2010Comput.Mater.Sci.}.
}
\label{fig:BZ_TET}
\end{figure}
To create a resonant cavity in an inverse woodpile photonic crystal, Woldering \textit{et al.} proposed a design whereby two proximal perpendicular pores have a smaller radius $(r' < r)$ than all other pores, as shown in Fig.~\ref{fig:crystalOfCavities}(a)~\cite{Woldering2014PRB}.
This structure has a tetragonal crystal symmetry, with the Brillouin zone and its high-symmetry path depicted in Fig.~\ref{fig:BZ_TET}.
Near the intersection region of the two smaller pores, the light is confined in all three directions to within a mode volume as small as $V_{\text{mode}}=\lambda^3$ where $\lambda$ is the free-space wavelength~\cite{Woldering2014PRB}. 
Supercell band structures reveal up to five resonances within the band gap of the perfect crystal, depending on the defect pore radius $r'$~\cite{Woldering2014PRB}. 
The best confinement occurs for a defect radius $r'/r=0.5$ that is also considered here. 

Fig.~\ref{fig:crystalOfCavities}(b) shows a 3D superlattice of cavities as is studied here, where each sphere indicates one cavity, as shown in Fig.~\ref{fig:crystalOfCavities}(a).
The cavity superlattice has lattice parameters ($c_s^{x}$, $a_s$, $c_s^{z}$) in the ($x, y, z$) directions that are integer multiples of the underlying inverse woodpile lattice parameters: $c_s^{x} = M_x c$, $a_s = M_y a$, $c_s^{z} = M_z c$. 
Here, we study the $M_x \times M_y \times M_z = 3\times3\times3$ superlattice such that the cavities are repeated every three unit cells with lattice parameters $c_s^{x} = 3c$, $a_s = 3a$, $c_s^{z} = 3c$. 
Thus, the cavity superlattice is also cubic, similar to the underlying inverse woodpile structure. 

We have calculated the band structure of the 3D cavity superlattice using the plane-wave expansion method \cite{Ashcroft1976book, Joannopoulos2008Book}, implemented in the recent version 1.5 of the MIT photonic bands (MPB) code~\cite{Johnson2001OE}. 
We employ the grid resolution $48\times68\times48$ per unit cell of the superlattice.
All calculations were performed on the ``Serendipity" cluster in the MACS group at the MESA$^{+}$ Institute.

\subsection{Optical setup}\label{sec:setup}

\begin{figure} 
    \centering
    \includegraphics[width = \columnwidth]{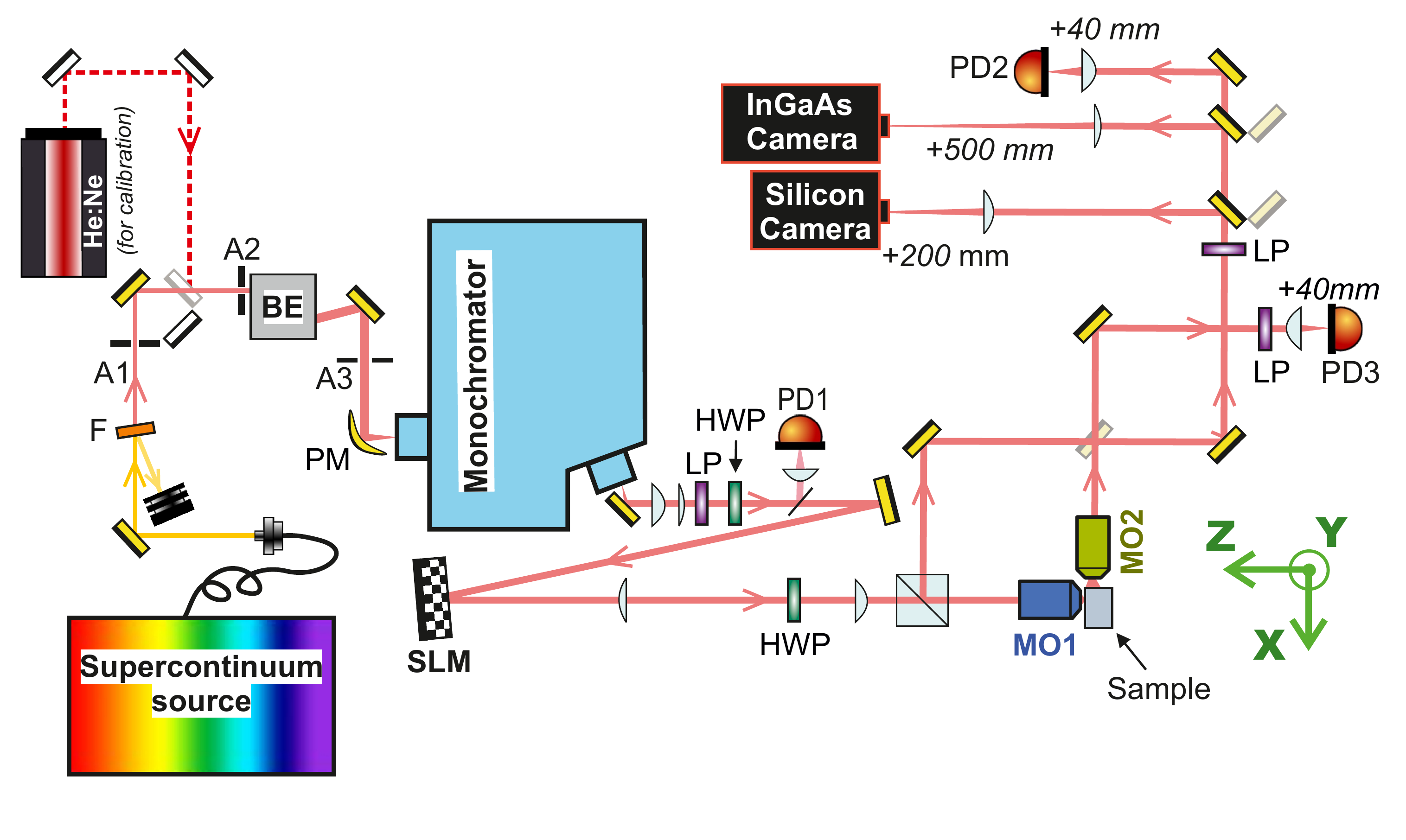}
    \caption{
    Optical setup to measure position-resolved microscopic broadband reflectivity, lateral scattering and wavefront shaping. 
    The Fianium SC is the broadband supercontinuum source, the long-pass glass filter F blocks the visible light at $\lambda < 850$ nm, the BE is a reflective beam expander. 
    The monochromator filters the light to a narrow band with a linewidth of about $0.7$ nm. 
    The linear polarization of the incident light is set by a half-wave plate (HWP, $\lambda/2$), and LP are linear polarizers to analyze the light. 
    The SLM is a spatial light modulator used for wavefront shaping and for correcting the direction of incoming light onto the sample. 
    Incident light is focused on the sample with a $100\times$ objective MO1 ($\mathrm{NA} = 0.85$) that also collects the reflected light. 
    Objective MO2 ($\mathrm{NA}=0.42$) collects light scattered from the sample in the perpendicular lateral (-X) direction that is detected by photodiode PD3. 
    The NIR camera (InGaAs) views the sample in reflection with an effective magnification of $250\times$. 
    The other Silicon camera can also be used to view the sample when using shorter wavelengths. 
    Photodiode PD1 monitors the incident light power, and PD2 and PD3 measure reflected and scattered signals from the crystal, respectively. 
    The sample is mounted on a 3-axis precision stage to adjust focusing and to focus input light to targeted positions. 
    The coordinates are shown on the sample, where incident light is along -Z direction and the lateral scattered light exits from the YZ surface. 
    }
    \label{fig:fullsetup}
\end{figure}

\begin{figure}[htbp]
    \centering
    \includegraphics[width = \columnwidth]{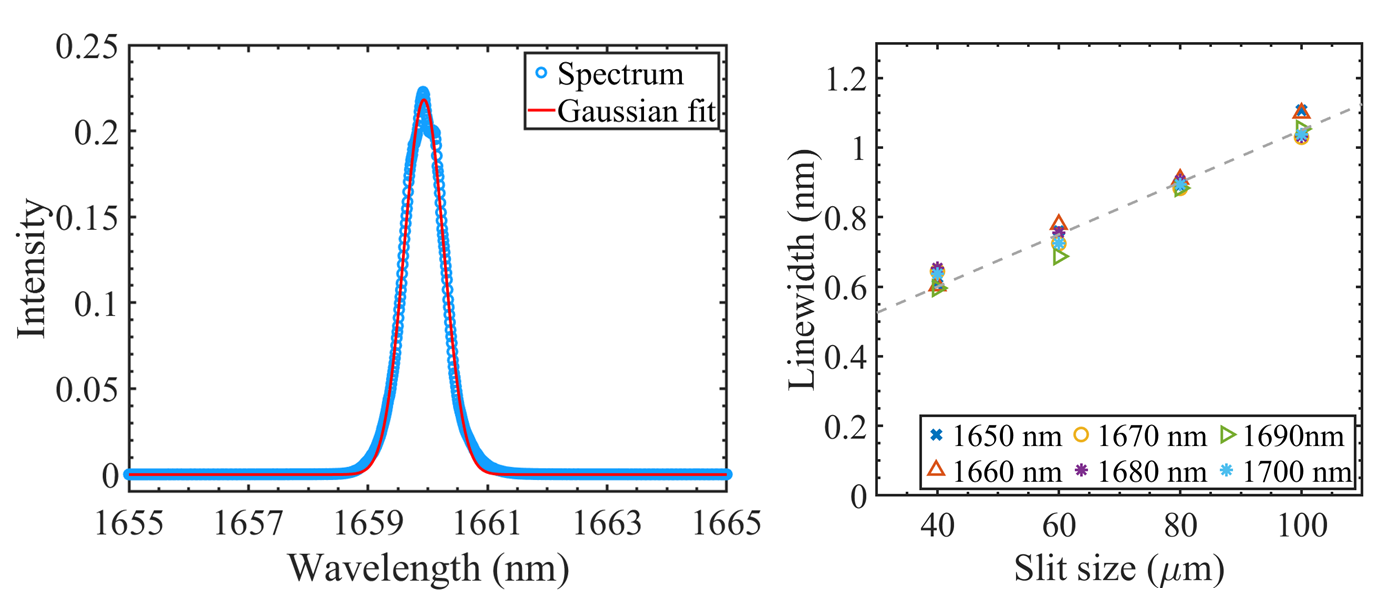}
    \caption{
    Left: The output spectrum of the monochromator for one single wavelength ($\lambda = 1660$ nm or $\omega_c = 6024~\rm{cm^{-1}}$) measured by a separate spectrometer.
    The linewidth is estimated by modeling with a Gaussian and taking the FWHM, yielding $0.78$ nm or $\delta\omega = 2.83~\rm{cm^{-1}}$, for a monochromator slit width of $60~\mu $m. 
    Right: The linewidth of the source versus slit width of the monochromator output at 6 different wavelengths, which is linearly proportional. 
    In all experiments, the width was set at $ 50~\mu $m.
}
    \label{fig:MC_output}
\end{figure}

We have developed a versatile near-infrared setup to collect position-resolved broadband reflectivity and lateral scattering spectra of photonic nanostructures, as well as to perform wavefront shaping to focus light inside samples. 
All measurements are processed in LabView environments. 
The near-infrared range of operation is compatible with 3D silicon nanophotonics as it avoids intrinsic silicon absorption. 
A pictorial representation of the full optical setup is shown in Fig.~\ref{fig:fullsetup}.
The setup consists of three main components: 
\begin{enumerate}
\item a broadband tunable coherent source,

\item a broadband wavefront shaper, and

\item twin-arm imaging of reflected and lateral (YZ-plane) scattered signals from the sample.
\end{enumerate}
The reflectivity part has been described earlier in Ref.~\cite{Adhikary2020OpEx}. 
The broadband tunable coherent source is realized by spectrally filtering the emission from a supercontinuum source (Fianium SC 450-4, 450 - 2400 nm) with a monochromator (Oriel MS257; 1200 lines/mm grating).
A long-pass filter (cut-off wavelength 850 nm) is used to reject the background from second-order diffraction of shorter wavelengths.
The filter is slightly tilted so that reflected light does not go back into the collimator and source (where it could lead to unwanted feedback and damage to the source), but gets absorbed in a beam dump. 
The filtered light beam is expanded in a reflecting beam expander (Thorlabs) and sent to a parabolic mirror (PM) to focus the light to fit through the input slit of the monochromator. 
The monochromator scans optical frequencies ranging from $ 4700 $ to $ 11000~\mathrm{cm^{-1}} $ (or wavelengths $900 < \lambda < 2120$ nm) with a linewidth of $0.6 \pm 0.1$ nm exiting from the output slit of width $50~\mu $m (see Fig.~\ref{fig:MC_output}), 
and a tuning precision better than $0.2$ nm. 
Since we use this setup also for wavefront shaping, we use sequential scanning of wavelengths instead of measuring the spectrum at once with a spectrometer as in~\cite{Thijssen1999PRL, Ctistis2010PRB, Huisman2011PRB}.
A He:Ne laser is used to calibrate the grating of the monochromator, and is not used for measurements.
The filtered light is then collimated and expanded to a beam diameter of $ 7.5 $ mm.
A small fraction ($8\%$) of the light is sent to a photodiode PD1 using a glass plate as a reference to monitor the input power.
The main beam is incident on a reflective phase-only spatial light modulator (Meadowlark optics; $ 1920 \times 1152 $ pixels; AR coated: $850 - 1650$ nm).
Since the output of the source is randomly polarized, a linear polarizer (LP) followed by a half-wave plate (HWP) are placed at the monochromator output to select the desired linear polarization orientation, since the polarization has to be parallel to the slow axis (in this case the Y-axis of the SLM) for the SLM to optimally function.
At this point, the setup is described in detail first from the point of view of spectroscopy and next wavefront shaping.

\subsection{Near-infrared reflectivity and lateral scattering}

In our optical setup shown in Fig.~\ref{fig:fullsetup}, the sample is mounted on an XYZ translation stage that has a step size of about $30$ nm. 
The reflected light from the SLM \footnote{For spectral measurements the SLM is used to reflect as a mirror.
In addition, the SLM is also used to correct a tilted beam in the setup as explained in the Appendix of this chapter.} is sent to an infrared apochromatic objective (Olympus LC Plan N 100$\times$) to focus the light onto the sample's surface with a numerical aperture NA $=0.85$.
The beam waist of the light reflected from the SLM is narrowed and re-collimated by a pair of lenses so that it fits the back aperture of the objective.
The diffraction limit in the setup is about $ 1~\mu  $m at longer wavelengths.
A second HWP is introduced on the beam path before the objective to rotate the linear polarization of the incident light on the sample for polarization-dependent measurements. 

Light reflected by the sample is collected by the same objective and a beam splitter directs the reflected light towards the detection arm where the reflection from the sample is imaged onto an IR camera (Photonic Science InGaAs).
In order to locate the focus of the input light on the surface, a NIR LED is used to illuminate the sample surface.
We use the XYZ translation stage to move the sample to focus the light on the desired location.
For example, an image of a 3D crystal as seen on the IR camera reveals the $XY$ surface of the Si beam, see for example Fig.~\ref{fig:speckleimage} in the main manuscript. 
The bright circular spot with a diameter of about 2 $\mu$m is the focus of light reflected from the crystal. 
The rectangular darker areas of about 8 $\mu$m $\times$10 $\mu$m are the XY surfaces of the 3D photonic crystals.
They appear dark compared to the surrounding silicon since the LED illumination is outside the band gap of these crystals whose effective refractive index is less than that of silicon.

Once the input light beam is focused on the sample, the reflected light is sent to photodiode PD2 (Thorlabs InGaAs DET10D/M, 900 nm - 2600 nm) by flipping out the mirror in front of the camera.
The photodiode records the reflected intensity $I_R$ as the monochromator scans the selected wavelength range.
At the same time, the scattered light from the sample in the lateral direction is collected by a long working distance apochromatic objective MO2 (NA = $0.42$), and sent to a third photodiode PD3 (similar to PD2).
An analyzer in front of each photodiode is used to select either parallel or cross polarization of the reflected and scattered light with respect to the input.
Thus, in total we get two sets of spectra as signal every time the wavelengths are scanned through the desired range: a reflectivity spectrum, and a lateral scattering spectrum. 
By moving the sample stage in the $\mathrm{Y}$-direction, the focus is scanned across the crystals with a step size of $\approx 0.4~\mu \mathrm{m}$, resulting in typically $30$ reflectivity and lateral scattering spectra per scan. 

\begin{figure}
\centering
\includegraphics[width=0.6\columnwidth]{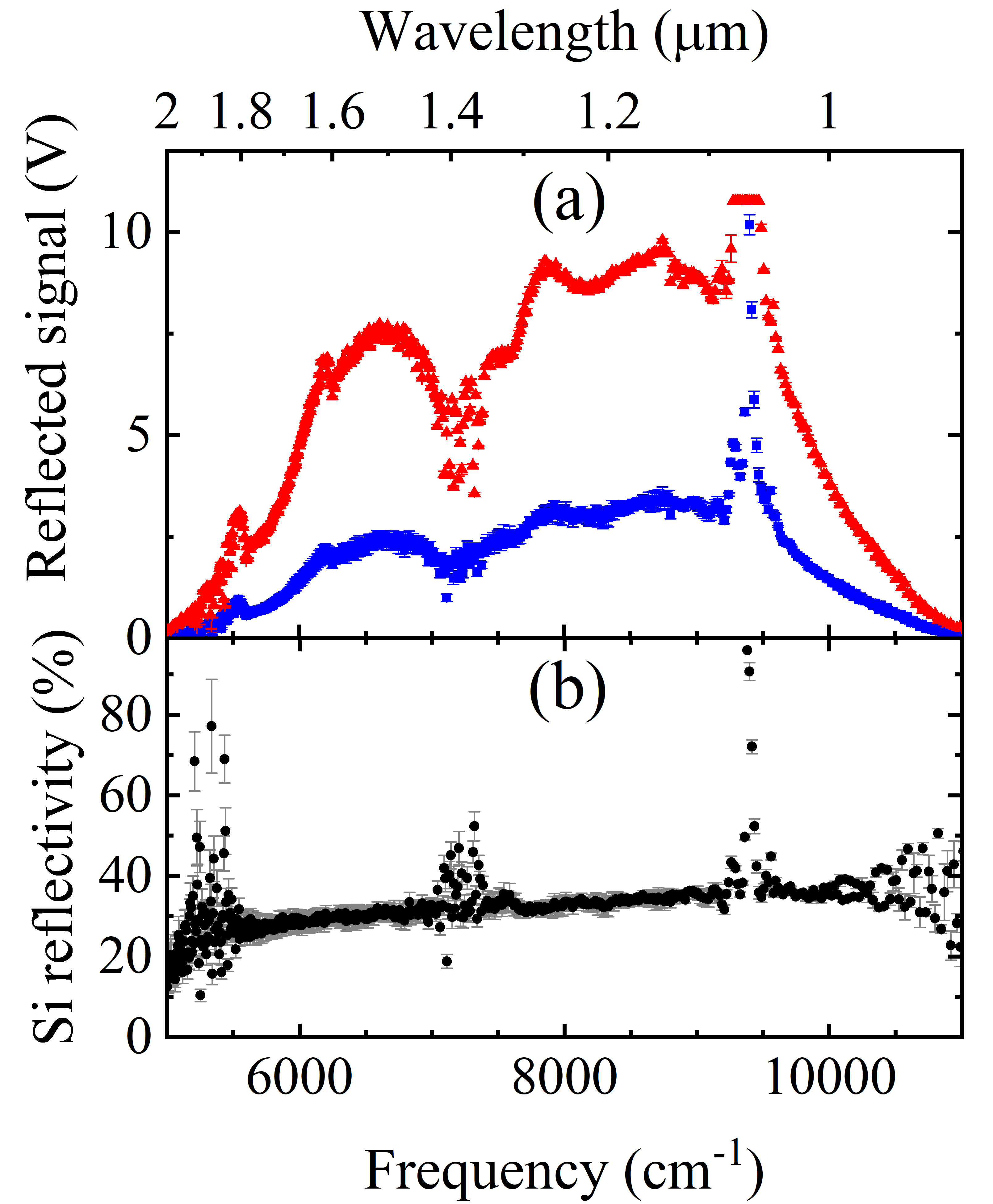}
\caption{ 
(a) Raw spectra recorded by photodiode PD2 of light reflected from a gold mirror (red triangles) and a clean silicon substrate next to the photonic crystals (blue squares). 
Due to the very high power near the pump frequency $9400~\rm{cm}^{-1}$ of the supercontinuum source, the intensity is usually saturated, hence this range is disregarded in our analysis of the measurements. 
(b) Reflectivity of bulk Si normalized to the gold reflectivity. 
}
\label{fig:ref_spectra}
\end{figure}
The raw reflectivity measured on two reference samples, \emph{viz.}, a clean gold mirror and the bulk silicon substrate are shown in Fig.~\ref{fig:ref_spectra}(a).
The spectra show that the intensity of the supercontinuum source varies markedly with frequency, hence proper referencing is important. 
The intensity cut-off at high frequencies is mainly due to limited detection bandwidth of the InGaAs photodiode and the cut-off at low frequencies is due to the absorption by the refractive optics (lenses, objectives, wave plates) and the limited bandwidth of the supercontinuum source. 
The range of noisy data near $7000~\rm{cm}^{-1}$ is attributed to water vapor absorption that may vary from day-to-day depending on lab conditions (and weather). 
Hence care is exerted while analyzing data in this wavelength range, by using data obtained in a narrow time window. 
In future, the noise may be suppressed by placing the setup in box purged with dry air or nitrogen. 
The intense peak at 1064 nm is the pump wavelength of the supercontinuum source; due to unavoidable detector saturation near this peak, data in this range is typically removed from analysis. 
Recording a typical reflectivity or lateral scattering spectrum takes about 5 to 25 minutes, depending on the chosen wavelength step size (typically 10 nm or 2 nm).
The calibrated reflectivity and lateral scattering are defined as
\begin{eqnarray}
R \equiv I_R/I_{0R}\\
LS \equiv I_{LS}/I_{0LS}
\end{eqnarray}
For reflectivity, the spectral response $I_R$ of the samples is referenced to the signal $I_{0R}$ reflected from a clean gold mirror that reflects $96\%$. 
Since it is tedious to dismount and realign the sample to take reference spectra during long measurements on crystals on a silicon beam, we also take secondary reference measurements on bulk silicon outside the crystals, which has a flat response $R \approx 31\%$ with respect to the gold mirror. 
The bulk silicon reflectivity is plotted in Fig. \ref{fig:ref_spectra}(b) that agrees with the expected $ \approx 31\% $ reflectivity of Si for normal incidence in the NIR spectral region~\cite{NSM}. 
The spectrum is flat \footnote{The Si reflectivity spectrum has a small positive slope with frequency since we do not consider the change of refractive index with frequency within this range.} without any features apart from the noisy data mentioned above.
Therefore, we use the Si signal also as a reference for reflectivity measurements. 

\begin{figure}
\centering
\includegraphics[width=0.6\columnwidth]{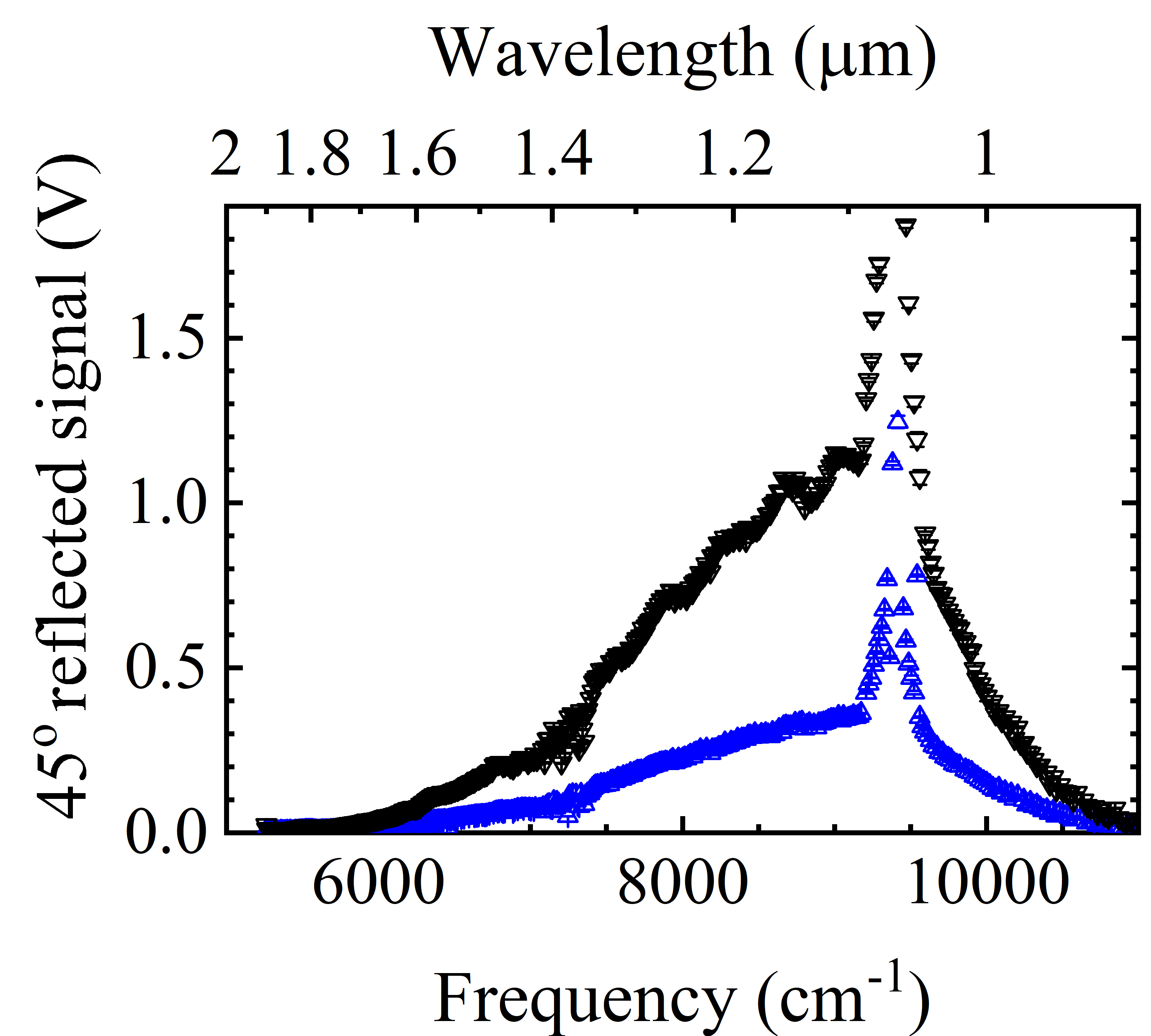}
\caption{
Raw spectra recorded by photodiode PD3 of light reflected at $45^\circ$ incident from a silicon surface for $s$-polarization (black downward triangles) and $p$-polarization (blue upward triangles). 
}
\label{fig:LSref_spectra}
\end{figure}

The reference for lateral scattering is a bit tricky since we need to have the input signal go through the objective MO2 for correct referencing.
Therefore we used the edge of a silicon beam that was cut at $ 45^{\rm{o}} $ to reflect the incoming light coming through MO1 towards MO2 and send it to PD3, just like the lateral scattering signal.
Using Fresnel reflectivity of $ 45^{\rm{o}} $ incident light on silicon, we use a calibration of $R_{p} = 43\% $ reflectivity for $ s $-polarized light and $ R_{p} = 19\% $ reflectivity for $ p $-polarized light.
To ensure that the signal to noise ratio of the photodiode response is sufficient to detect signal in the desired range, each detector photodiodes are fed into a lock-in amplifier to amplify the signal with a suitable gain. 
Since a serial measurement mode holds the risk of possible temporal variations in the supercontinuum source, we simultaneously collect the output of the monochromator with photodiode PD1 in each reflectivity scan.
An example of reference spectra for LS measurements is shown in Fig. \ref{fig:LSref_spectra} for two orthogonal polarizations of the incident light.
All measurements described above are controlled by and the data saved by a home-built LabVIEW program, that also allowed for remote control, which turned out to be very convenient when access to our laboratories was restricted during the Covid pandemic.

\begin{figure}[htbp]
    \centering
    \includegraphics[width = 0.5\columnwidth]{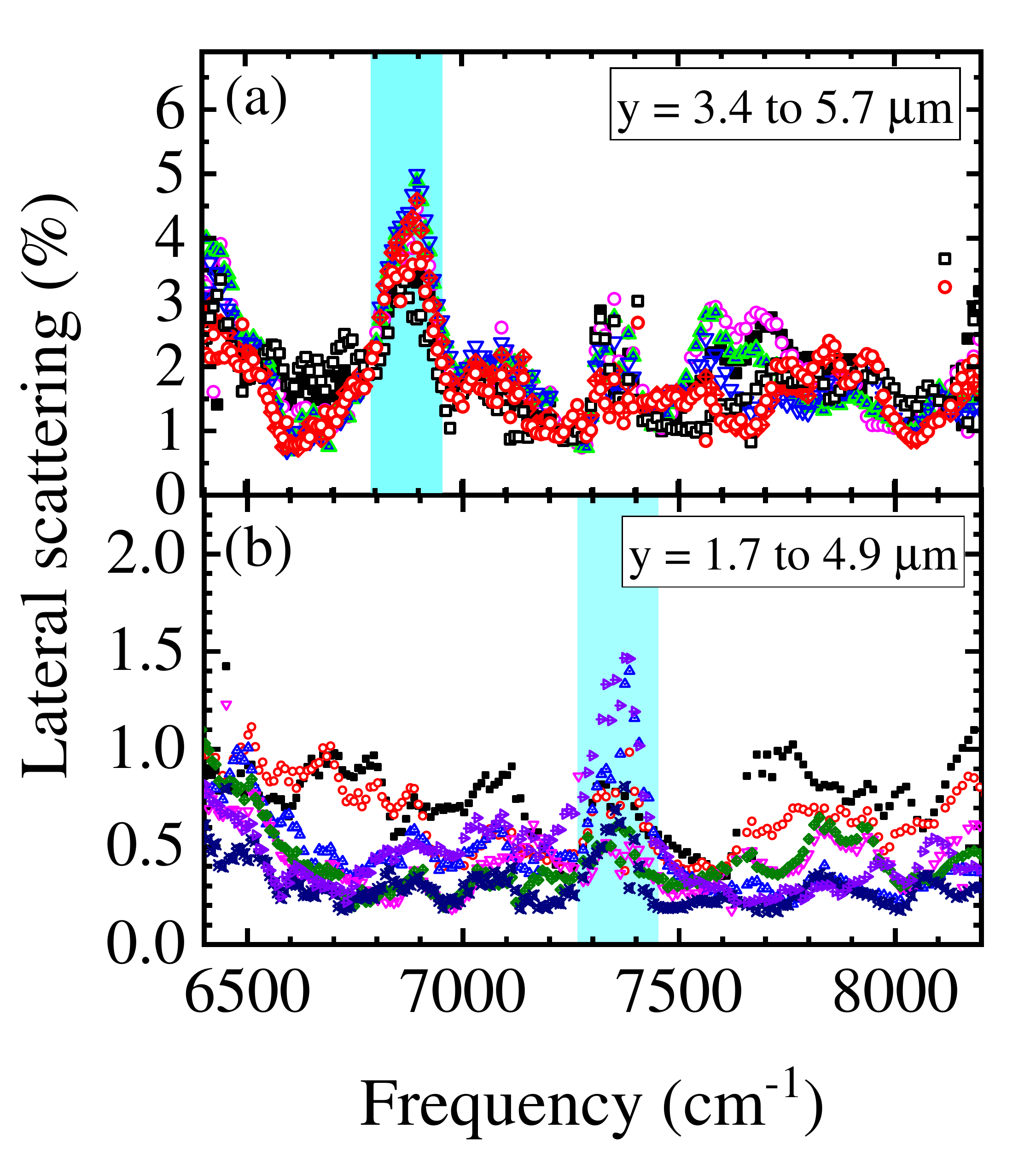}
    \caption{
    (a) Multiple lateral scattering spectra collected on superlattice SL5 at different y-positions in the vicinity of a defect show that the peak near $6900~\mathrm{cm^{-1}}$ is reproducible, whereas other peaks do not reproduce and are hence identified as random speckle. 
    (b) Multiple spectra across three defect pores, show a reproducible peak near $7386~\mathrm{cm^{-1}}$ for superlattice SL3.
    The Y-positions of the spectra are given in the legends while the $ X $-position is nearly constant (about $2.5 \mu$m). 
    The frequency range shown covers roughly the original band gap of the unperturbed crystal. 
    }
    \label{fig:LSspectra}
\end{figure}

\subsection{Distinguishing superlattice scattering from random speckle}

To distinguish a superlattice peak from random speckle peaks~\cite{Goodman2007book,Dainty2013book} that arise as a result of random and unavoidable structural variations~\cite{Koenderink2005PRB}, we monitor the reproducibility of the peaks while scanning the illuminating spot across the sample surface. 
Therefore, we plot multiple lateral scattering spectra at different $Y$-positions near the defect pore (at $X = 2.5~\mu$m), as shown in Fig.~\ref{fig:LSspectra}(a) and (b) for the superlattices SL5 and SL3, respectively. 
It is clearly apparent that the peak near $6900~\mathrm{cm^{-1}}$ reproduces at all positions near the defect pore, with gradually decreasing intensity as the focus moves away from the defect pore. 
We also observe a second reproducing peak near $7386~\mathrm{cm^{-1}}$, possibly a second superlattice band.
For superlattice SL3 (Fig.~\ref{fig:LSspectra}(b)), we also see a scattering peak near $7386~\mathrm{cm^{-1}}$ that reproduces even across multiple defect pores. 
All other scattering peaks do not reproduce, hence they are uncorrelated speckle arising from unavoidable disorder of the crystal. 
Therefore, we conclude that all reproducing peaks are spectral features of the intentional superlattice of defects.

\end{document}